\documentclass[MSNbibl,number,citesort,dvips]{arxstspdf}
\usepackage{flushend}
\usepackage{stfloats}
\usepackage{graphicx}
\volume{28}
\issue{2}
\pubyear{2013}
\firstpage{269}
\lastpage{281}
\doi{10.1214/12-STS404} %kopijuoti is PTS
\begin{document}
\begin{frontmatter}

\title{Another Conversation with Persi Diaconis}
\runtitle{Another Conversation with Persi Diaconis}
% kai straipsnis turi susijusiu diskusiju ir rejoinder'iu
%rejoinder at \relateddoi{r}{10.1214/00-STSXXXX}.}

\begin{aug}
\author{\fnms{David} \snm{Aldous}\corref{}\ead[label=e1]{aldous@stat.berkeley.edu}\ead[label=u1,url]{www.stat.berkeley.edu/users/\textasciitilde aldous}}
\runauthor{D. Aldous}

\affiliation{U.C. Berkeley}

\address{David Aldous is Professor, Statistics Department, University of California, Berkeley, 367 Evans Hall,
Berkeley, California 94720, USA \printead{e1,u1}.}

\end{aug}

% ABSTRACT
%
\begin{abstract}
Persi Diaconis was born in New York on January 31, 1945.
Upon receiving a Ph.D. from Harvard in 1974 he was appointed Assistant
Professor at Stanford. Following periods as Professor at Harvard
(1987--1997) and Cornell (1996--1998), he has been Professor in the
Departments of Mathematics and Statistics at Stanford since 1998.
He is a member of the National Academy of Sciences, a past
President of the IMS and has received honorary doctorates from Chicago
and four other universities.

The following conversation took place at his office and at Aldous's
home in early 2012.
\end{abstract}

% KEYWORDS
\vspace*{3pt}

\end{frontmatter}

%s1 #&#
\section{Markov Chains, Mixing Times and Monte Carlo}

\textbf{Aldous:} You were interviewed in October 1984 for a \textit{Statistical
Science} conversation article~\cite{MR858513}, so I~won't ask about your
earlier personal and academic life, but try to pick up from that point.
You and I~were both involved, in the early 1980s, with the start of the
topic now often labeled ``Markov chains and mixing times''~\cite{MR2466937}.
Can you tell us your recollections of early days, and give some overview
of how the whole topic has developed over the last 30
years?\looseness=-1

\textbf{Diaconis:} That's been the main focus of my work since the 1980s,
and it started for me with an applied problem. I~was working at Bell
Labs and we were simulating optimal strategies in various games and
needed a lot
of random permutations. The standard way is to pick a random number
between $1$ and $n$ and switch it with $1$, then pick a random number
between $2$ and $n$ and switch it with $2$, etc. If you do that
$n-1$ times, it's exactly random. We got the results of many hours
of CPU time of simulations and something looked wrong. There were 2000
lines of code and we looked for a mistake. After three days we asked,
``How did you generate the random permutations?'' The lady said,
``You said that fussy thing but I~made it more random---I~switched a
random card with another random card.'' I~said, ``You have to redo the
simulations'' and she said, ``It's crazy.'' Then she went to her boss
and he went to his boss and he came down and yelled at me. ``You
mathematicians are crazy---she did a hundred transpositions and that
has to be enough
with 52 cards.'' So I~really wanted to know the answer---how many
transpositions does it take to mix 52 cards? I~came back to the West
Coast in the early 1980s, talked with people like you and Rick Durrett
and we saw that if you did it for order $n^2$ times that would be
enough \ldots

\textbf{Aldous:} \ldots by an easy coupling argument \ldots

\textbf{Diaconis:} \ldots but it wasn't clear if that was the right answer.
Eventually Mehrdad Shahshahani was here and we realized we could
set it up as a problem in Fourier analysis and carefully do the Fourier
analysis on a noncommutative group and get the right answer and it
turned out~\cite{MR626813} to be $\frac{1}{2} n \log n$, which when $n
= 52$ gives $103$.
That was for me the start of it. And in 1983 you wrote this article
\cite{MR770418} on mixing times for Markov chains.
Around the same time Jim Reeds had become interested in riffle shuffling.
He had reinvented a model that Gilbert and Shannon had invented and had
numerical results and ideas but couldn't seem to push them through.
You and I~started to talk about it and invented stopping time
arguments~\cite{MR841111} that turned out to give good answers in some
cases. Spurred on by these two examples I~started to think hard about
mixing times.
Around the same time, what I~now call ``the Markov chain Monte Carlo
revolution''~\cite{MR2476411} began with the paper by Geman and Geman
\cite{geman}. So the topic of mixing times became about more than just
card shuffling, it was also about how long should you run a simulation
until it converges.
I~say now, as I~said then, that if you take any application of MCMC in
a real problem and ask if we theoreticians can give a sensible answer
to a practitioner about
``how long \ldots,'' then we can't. These are open research problems,
every one of them.
We have ideas, we have heuristics, but as math problems they are really open.

A fitting proof of that is the following:
The Metro\-polis algorithm, Glauber dynamics, the Gibbs sampler and
molecular dynamics
were all invented to
solve one problem, the problem of random
placement of hard discs in a box. Take, say, two-dimen\-sional discs of
radius $\varepsilon$ in the unit square. You want them to be uniform
random subject to nonoverlap. The Metropolis algorithm is that you pick
a disc at
random, you try to move it a little, if it's possible to move then do
it, if not then try another pick. Glauber dynamics is similar. But as
far as I~know---despite billions of steps of simulations over the last
60 years---nobody has ever sampled from anything close to the
stationary distribution, in the interesting case of high disc density.
There's supposed to be a phase transition around 81\%, but the
algorithms have no hope of converging near that point and yet people
get numbers
from the simulations and talk about them. I~think the same goes for statistical
algorithms too---people who don't want to think about it just run the
simulation until something seems to have settled down. So I~think the
current state of the art is there's a ton of research still to be done;
everyone finds us theoreticians annoying prigs for asking what can you
show rigorously. But it's not just being annoying. In enough cases the
algorithms really don't converge, and people don't seem to want to own
up to that.
In~\cite{MR2504885,MR2819161} we tried pretty seriously to do the hard
discs problem, but there's still a very long way to go.

\textbf{Aldous:} Now there's a distinction between ``don't converge'' and
``we can't prove they do converge'' \ldots
And there's an argument that in practice one uses ``black box'' methods
like MCMC in complicated situations where you don't have any nice
structure, whereas to do any theoretical analysis you need to assume some
structure, so we (perhaps) are in a Catch-22 situation where one can do
theory for MCMC only in situations where you wouldn't actually use it.
And then you have to rely on the heuristics that applied researchers
have developed.

\textbf{Diaconis:} Well at least for the chemists I~talk to, who study
molecular dynamics, they haven't converged,
they're in some kind of local minimum, and really dramatically new
ideas are needed.
I~think a very interesting research question is to
look at the zoo of diagnostic techniques that are available today,
and look at the hundreds of examples of Markov chains about which we
know everything. Take some of those examples and diagnostic techniques
and try to see how they behave. That seems like a very reasonable thing
to do. I've tried for 20 years to get a graduate student to do this,
but somehow I~can't get anyone to sit down and do the work. I~should
have learned I~need to do it myself. They're hard math problems. The
diagnostics can be pretty sophisticated; they're not just second eigenvalue
%diagnostics}
but involve \textit{sups} and \textit{infs} of complicated functionals.
We do have a lot of machinery and they're nice math problems
so this project would be useful to help evaluate diagnostics.
What's annoying to me is how little that problem is recognized.
If you go to a Statistics meeting, in talk after talk somebody runs the
Gibbs sampler because that's the standard thing to do, and they say
they ran it 10,000 times and it seemed to be OK, and they just go on
with what they're doing. People don't even try to prove the chain
does what it's supposed to do, that is, have the desired stationary
distribution.

%s2 #&#
\section{Bayesian Statistics}
\textbf{Aldous:} Let me move on to Bayesian statistics, which has been
a recurrent
feature of your research. Maybe I~should remind readers that 30 years ago
this was completely unrelated to Markov chains, but now a major use of
MCMC is for computing Bayes posteriors---but let's leave MCMC for later.
I'm not competent to ask good questions here, so let me just throw out
two points and then I'll sit back and listen.

(a) You have work, such as the 1986 paper~\cite{MR829555}
with David Freedman, that addresses foundational technical
(rather than philosophical) issues in Bayes\-ian statistics.

(b) There is a recent newsletter piece by Mike Jordan~\cite{jordan}
summarizing comments
by many leading Bayesian statisticians (including you) on open problems in
Bayesian statistics.

So I~guess I~am asking for your thoughts on the history and current state
of methodological/techni\-cal aspects of Bayesian statistics.

\textbf{Diaconis:} I~came into Statistics late in life, becoming aware
of the Bayesian position when I~was a graduate student at Harvard.
Art Dempster and Fred Mosteller were Bayesians---not everyone,
Bill Cochran wouldn't dream of doing anything Bayesian.
I~read de Finetti's work and found it frustrating and fascinating, as I~still do today,
but it was inspiring and so I~tried to make my own
sense out of it.
One of the things I~noticed was that de Finetti's theorem involves an infinite
exchangeable sequence. I~wondered whether there could be a
finite version, saying the sequence is almost a mixture of IIDs.
In fact, I~wrote my first paper on that topic when I~was
still a graduate student~\cite{MR0517222}. That was when I~came to meet
the Berkeley
people---David Freedman and Lester Dubins and David Blackwell,
the latter two being Bayesians of various stripes---and, in fact,
David Freedman's thesis (published as~\cite{MR0137156}) was about de
Finetti's theorem for Markov chains.
The story, in a nutshell, is that David was a
precocious but difficult young man who wanted to do a thesis in
Probability, and (the way David
told it) he went into Feller's office and Feller looked up, said,
``prove de Finetti's theorem for Markov chains,'' looked back down, and
David left. So he went and did it. In order to prove the theorem, he
needed to assume the chain was stationary. When I~met him
at a Berkeley-Stanford joint colloquium, I~said that I~knew how to do
it without stationarity. I~could make a finite version of the theorem
and it didn't need anything like stationarity. He agreed to listen, and
Lester did too. Lester was very dismissive, but David wasn't, and that
led to our work on finite versions of both the Markov and the IID cases
\cite{MR556418,MR577313}.

I've written far too many papers.
I'll try to distinguish the ones that people seem to like into
Statistics or Probability or something in between.
You presented me with a list of papers \ldots

\textbf{Aldous:} your 30 most cited papers, according to Google Scholar,

\textbf{Diaconis:} \ldots and about a third are Statistics and a third
are Probability
and a third are in between, like de Finetti's theorem, which I~was
interested in for
philosophical reasons, trying to make sense of the way model-building
goes. I~like de Finetti's take, focusing on observables, and I'd like
to understand just what you need to assume about a process, in terms of
observables, in order for it to be a mixture of standard parametric families,
a mixture of exponential or normals or some other thing.
That led to a lot of work~\cite{MR898502}.
That era seems to have quieted way down---nowadays no one works
on exchangeability particularly, though a few of us still dabble in
it.\looseness=1

About a year ago, some of our chemists here came to me. They were
working on a
protein folding problem with the IBM Blue Gene project.
They're really doing protein folding---taking forty molecules and
ten thousand water molecules and then doing the molecular dynamics
to see how the protein folds by using the equations of physics.
It's a very high-dimensional system---one particle is represented by
twelve numbers---and the chemists were coarse-\break graining and dividing this
high-dimensional space into maybe five thousand boxes. Their hope is
that within a box it will quickly get random---in the sense of
invariant measure for a dynamical system---and that jumps from box to
box can be modeled as some Markov chain. Refreshingly to me, they were
Bayesians, so
they wanted to put a prior on transition matrices and, because the laws
of physics are reversible, they wanted the prior to live on reversible chains.
I~realized that some earlier work with Silke Rolles~\cite{MR2278358}
exactly gave the conjugate prior for reversible Markov chains.
I~told them about it, they implemented it and they say it makes a big
difference.
There's a marvelous graduate student here, Sergio Bacallado, he's a chemist,
and he's written papers such as~\cite{MR2816340} in the Annals
which extends our work on priors in more practical directions.
There's something very exciting here---our old work had horrible
formulas involving quotients of Gamma functions and now someone is
caring to get it right, and thinking it's sensible. So that subject is
quite alive and well today,
although Sergio has taken it a lot further.
One of the main problems for Markov chain theory is to make the mixing
time theory for continuous-space chains. There really are technical
difficulties for continuous spaces, and he's managed to get around that.

Now in a larger view, it's a very exciting time for Bayesian statistics.
When I~first learned about it, in the early 1970s, it was still Good
and Savage, and people were still arguing about whether an egg in a
fridge is rotten or not \ldots

\textbf{Aldous:} \ldots and the Bayesian lady tasting tea.

\textbf{Diaconis:} I~remember going to my first Valencia meeting.
One of the world's leading Bayesians, John Pratt, a marvelous man,
was analyzing some data, his wife's estimates of upcoming gross
receipts at a cinema where she worked in Cambridge, MA. He was doing regression,
and at the end he did an ordinary least squares but nothing Bayesian
\cite{pratt}. I~asked him why not Bayesian?
He said it was too hard to figure out the priors and it wouldn't have
made any
difference anyway. I~was shocked and dumbfounded. That was 1983, but
since then we can actually implement Bayesian methods. And we do.
Now the judgement has to be put off---frequentist methods have had 200
years of people tinkering with them and we're just starting to use
Bayesian methods. I~think it's reasonable to let time settle down before
deciding whether they are better or worse. There are lots and lots of
groups doing Bayesian analysis.

One of the big tensions in Statistics, which is a mystery to me, is
really big data sets.
You can try to estimate huge numbers of parameters
with very few data points. Now I~understand sometimes there's a story
that seeks to justify that, but it makes me very, very nervous.
If you try to think about being a Bayesian in that kind of
problem, it can't be that you have any idea about what priors you're
putting on,
you're completely making something up. It's nothing other than a way of
suggesting procedures. It might be useful, it might not be useful.
There are a lot of people trying to do that, but it's a completely
different part of the world and I~don't have much feeling for it.
It's so taken over Statistics right at the moment that I~feel compelled
to put in the following sentence. There \textit{are} huge data sets;
there \textit{are} also
many, many small data sets. And that's where the inferential
subtleties matter. If you're sick and you're trying to think about a new
procedure for your tooth and there are two available procedures,
with 10 or 50 instances of each \ldots
what should you do?
Statistics encounters lots of problems like this too.
So it's good to remember that while there are huge data sets and that's
very exciting, there are also lots of small data sets and there's still
room for the classical way of thinking about statistical problems.

\textbf{Aldous:} A cynical view is that there's more money in the
fields with
big data sets.

\textbf{Diaconis:} Tsk tsk (laughs), you won't get any argument from me.

%s3 #&#
\section{Teaching the Philosophical Foundations}

\textbf{Aldous:} You teach an undergraduate course with Brian Skyrms on the
philosophical foundations of Statistics. You describe its
topic as ``10 great ideas about Chance.'' Now most readers of
\textit{Statistical Science} have surely never taken, let alone taught,
such a course. Can you
tell us about the course?

\textbf{Diaconis:} Philosophers and statisticians have\break thought for a very
long time about what probabilistic statements mean and how to combine
disparate sources of information to reach a conclusion. These are still
important questions and not ones to which we know the answer. We begin
our course with the first great idea, that probability can be
measured---the emergence
of equally likely cases, the first probability calculations.
There is of course a discussion of frequentism and of various kinds of
Bayesians. Indeed, I.J. Good once wrote an article
entitled 46656 \textit{varieties of Bayesians}
where he states 11 ``facets'' like whether utility is emphasized or avoided,
whether physical probabilities are denied or allowed, and so on. We try
to explain some of the different kinds of Bayesians.
Brian and I~are both subjectivists---I~am what I~call a
nonreligious Bayesian, that is, \textit{I} find it useful and
interesting and I~don't really care what \textit{you} do. Some of the
course is pointing out the shallowness of naive frequentism.
Bayesians are happy to talk about frequencies, in that when you have a
lot of data the data swamps the prior, and you will use the frequency
in order to make your inferences. It's not that Bayesians argue against
frequencies, they're happy to have a lot of data, and frequencies are
forced on you by the mathematics.
So we discuss and prove those things. We also explain von Mises
collectives, which have morphed into the complexity
approach to probability.\looseness=-1

One of the things I~find interesting that's hard to make philosophy out of,
is what I~want to call the von Mises pragmatic approach.
If you ask working statisticians what they think probability is,
they say, well,
you do something a lot of times, and it's the proportion of times
something happens.
If you ask about the probability Obama will be re-elected,
they will respond with a cloud of words.
Or they'll walk away or say it's too difficult to talk about.
What von Mises said is that any scientific area has practice and theory.
He discusses geometry---there's the mathematical notion of circles and
straight lines, then there's
practical architecture and drawing. The theory can be used, but at some
point you have to relate\vadjust{\goodbreak} the theory to the real world.
I~think that sort of pragmatic approach to foundations is important.
But von Mises never tells you how to do so.
I~ask this question for differential equations.
If some guy writes down a differential equation,
and there's a picture of water whirling around in a vessel
with blockages---what does that equation have to do with the whirling of
the water? In order to answer that, many of us would say,
``That's what Statistics is about.'' Whether theory fits data is a
statistical question. So we can apply this to our own subject: does
statistical theory fit the real world?

Anyway, we hope to turn the course into a book, after several years of
iterations.

\textbf{Aldous:} What kind of students take the course?

\textbf{Diaconis:} About 70 students, undergraduates or graduates
in Statistics or Philosophy, and just interested other people, even
some faculty
attend. It's quite lively, there's lots of discussion. We teach it once
a week
for three hours, which is exhausting for everyone concerned.

Trying to think about why we do what we do is important,
but nobody talks about it.
I~tell the following two stories. One is about you, and one is about
Brad Efron. At some stage you and I~were talking, as we often do, and I~said
I~was going to teach a course on the Philosophy of Probability.
And you got quite irate, saying, ``You're just going to tell a bunch of words
that won't illuminate anything.'' And my good friend Efron got
similarly very angry. He said, ``That's just going to be that Bayesian
garbage,'' reached into his pocket, took out a handful of coins, threw
them, and said, ``Look: Head, Tail, Tail, \ldots---\textit{that's} random.''
So people hear ``Philosophy'' and take it in a religious way.
To me, the question
``is what you're doing really \textit{about} anything?'' is worth discussing,
and we're just trying to talk about it.

If you want to know what the problems in Bayesian statistics are, ask a
Bayesian. We know! It's very hard to put meaningful priors on
high-dimensional real problems. And the choices can really make a
difference. I'm going to give one example of that, just for fun.
Suppose you're teaching an elementary Probability course. It's the
first day of term, you walk into class, you see there are 26 students
in the class,
so you decide to do the birthday problem.
Here are two thoughts about the birthday problem.
First, if it doesn't work, then it's a disastrous way to start a course.
Second, the usual calculation assumes each day is equally likely.
But my students are about the same age, and there are more births on
weekdays than weekend-days---that's about a 20\% effect---and then
there are smaller seasonal effects. So the uniformity might not be true
for my class. We don't really know what the probabilities are. So let
me put a prior on $(p_1,p_2,\ldots,p_{365})$. If your prior is uniform
on the simplex,
then the key number of people (to have a 50\% chance of some birthday
coincidence) decreases from 26 to about 18.
For the coupon collector's problem,
using a story that Feller suggested, the key number of people in a village
(to have a 50\% chance that every day is someone's birthday) is about 2300.
That's under the uniform multinomial model. If instead you take the
uniform prior on the simplex,
then---it's a slightly harder calculation to do---but if I~remember,
the key number increases to about 190,000. That's a little surprising
when you first hear it, but under the uniform prior some $p_i$ will be around
$(1/365)^2$ so you need order $365^2$ people just to have a good chance
of having that one day as a birthday.

\textbf{Aldous:} But isn't this a good argument against the naive
Bayesian idea
of inventing priors that are mathematically simple but without any
real-world reason?

\textbf{Diaconis:} Sure, and that was the point of the exercise.
Bayesian statisticians \textit{should} be thinking more carefully about
their priors. Part of that is understanding the effect of different
priors, and those are math problems. In the birthday problem,
math\break
showed the prior didn't have too much effect, whereas for the coupon
collector's problem it had a huge effect. Susan Holmes and I~wrote a
paper called \textit{A Bayesian peek into Feller volume} I \cite
{MR1981513} taking his elementary problems and making Bayesian versions
of them. When does it make a difference and when not?
It's a paper I~like a lot.

\textbf{Aldous:} A version of the nonuniform birthday problem I~give in
my own ``probability and the real world'' course~\cite{real-world} is
to take
$p_i = 1.5 \times\frac{1}{365}$ for half the days and $0.5 \times\frac
{1}{365}$\vspace*{1pt} for the other half. This makes surprisingly little
difference---the key number decreases from 23 to 22. And to avoid the
possible disaster of it failing with my students, I~show the
active roster of a baseball team
(easily found online; each MLB team has a page in the same format)
which conveniently has 25 players and their birth dates.
The predicted chance of a birthday coincidence is about 57\%.
With 30 MLB teams one expects around 17 teams to have the coincidence;
and one can readily check this prediction in class in a minute or so
(print out the 30 pages and distribute among students).\looseness=1

%s4 #&#
\section{Books: On Magic and on Coincidences}

\textbf{Aldous:}
On a lighter note, I~have found myself following in your footsteps in various
aspects of academic life, a minor such aspect being ``unfinished books.''
The 1984 conversation refers to the book on coincidences you were writing
with
Mosteller, and there is a 1989 joint paper~\cite{MR1134485}, but when
can we
expect to see the book?

\textbf{Diaconis:}
Well, there were two books mentioned in that interview, and the \textit
{other} one, with Ron Graham on mathematics and magic, has recently been
published~\cite{magic}. So it took 27 years, but we did finish it.
I'm starting to think about the coincidences book again.
We're sitting in my office and you see those folders up there \ldots

\textbf{Aldous:} I~see about 15 of those very wide old open-ended cardboard
files \ldots

\textbf{Diaconis:} \ldots those folders have newspaper clippings
collected by Fred Mosteller over 30 years, and every one has a few
pages saying
here's a kind of coincidence we might study via a model, and here's
some back-of-an-envelope calculation.
I~give a lot of public talks, about 50 a year, and I~had stopped giving
the talk on coincidences, but I've now committed to giving the talk
again in a few weeks. That's how I~trick myself and get back into
thinking about the topic.
So look for the book sometime in the next five years.
I~promised Fred (before he died in 2006) I~would do it, and I'm going
to gear up and do it.\looseness=1

\textbf{Aldous:} The colorful story of you running away from home at
age 14 to do
magic, then buying Feller and teaching yourself enough mathematics to
understand it, was told in the 1984 interview, and has become well known
in our community. But I've joked to students ``if you meet the Queen of
England, don't slap her on the back; if you meet Persi Diaconis, don't
ask him to do a magic trick.'' Now that you and Ron Graham have
published the book on mathematics and magic~\cite{magic}, could you
tell us
a little about what's in the book?

\textbf{Diaconis:} The reason I~first got interested in mathematics was
via magic. I~had hoped to call the book \textit{Mathematics to Magic
and Back}, but the publisher vetoed that, saying people wouldn't get
the idea.
Now it's called
\textit{Magical Mathematics: The Mathematical Ideas that Animate Great
Magic Tricks}, maybe a bit pompous. One of the things about mathematics
and magic is that if some person says,\break ``I~know a card trick,'' you
wince inside, because they're going to deal cards into piles\vadjust{\goodbreak} on the
table, and everyone's going to fall asleep. How long until I~can change
the conversation?
We're interested in \textit{good} magic tricks, which are performable and
don't look mathematical, but which have some math behind them. Some of
the math turns out to be pretty interesting.
Most of the tricks are ones we invented ourselves, which is why we don't
get strung up for revealing secrets; the magic community doesn't like
that, but we seem to get away with it. There's not much probability in
the book---there's some material on riffle shuffles and that sort of
thing---and some old tricks of Charles Jordan that we made mathematical
sense of.
To whet your appetite, there's a chapter on the connection between
riffle shuffles and the Mandlebrot set.

\textbf{Aldous:} Science has a notion of progress---one\break
could take any scientific topic and write a nontechnical article
on progress in that topic over the last 30 years. Is there an analog of
progress in magic?

\textbf{Diaconis:} Here I'm a bit negative. The final chapters are about
who are the current stars---who is inventing tricks that are new and
really different? The people we describe are old or now departed.
The younger people don't seem to be inventing math-based tricks.
But in the coming quarter I'll be teaching a course on mathematics and
magic here at Stanford, so I'm trying to cultivate young people myself.
Magic is changing in many ways, and the main one is again negative.
Because of Wikipedia and youtube there are very few secrets any more.
You could be watching a show and type the right words into your
smartphone and get an explanation, and this won't go away.
It's profoundly changing magic, likely not for the better.

Now I~do have a positive hope---maybe this will encourage people to
invent new and better tricks. Also \ldots
when I~was a kid, I~was once hanging around with my magic mentor Dai Vernon
at a billiard parlor. Billiards is a very refined game, the gentleman's
version of pool. Now pool halls are notoriously rowdy, smoke-filled
with gambling and drinking. This was a group of people, seated around
two masters,
playing three-cushion billiards. The crowd was silent aside from an
occasional quiet \textit{ooh} of appreciation.
Vernon looked at me and said, ``Wouldn't it be wonderful if people
watched magic that way.'' If people would learn a bit more about magic and
appreciate the skill and presentation, then maybe it would become like
watching a classical violinist. Those are my dreams about how exposure
might change magic for the better.

%s5 #&#
\section{Collaboration with David Freedman}
\textbf{Aldous:} We've already mentioned David Freedman, my long-time
colleague at
Berkeley, and perhaps your major collaborator, who sadly died in 2008. I~regarded
him as one of the handful of people in our business who are
unique---there was nobody \textit{like} David. I~mentally pictured him as
Mycroft Holmes (Sherlock's smarter older brother, who appears briefly in
several stories to give sage advice) and I~recall you having some
``bright light'' image. Can you tell us some
things about your collaborations and about David's impact on the field?

\textbf{Diaconis:}
I~first met David at a Berkeley-Stanford joint colloquium barbeque
at Tom Cover's house. I~had read his thesis when I~was a graduate student,
so I~had something to say to him. He was a very crusty character. He
had a kind of ``gee shucks, I'm just a farm boy'' outer style, but he
was in fact the debating champion of Canada. He was an honest man, and
there aren't so
many of \textit{them}. He could be difficult.
There's an image---that I~heard from Jim Pitman who maybe heard it
from Lester Dubins---of David working on a problem: you'd ask him a
question and he would berate you and say that's stupid, but then he
would get down and focus. And when he was focused it was like there was
this very bright clear
light on a narrow part of the problem, and then it would shift slightly over
and focus on a next part. That was how he worked. He wasn't a quick
glib guy.

At some stage he decided that the main impact he could make in Statistics
was what he called \textit{defensive statistics,} which was trying to make
an art and science out of critiquing knee-jerk modeling and the wild
misuse of probability models. He was as effective as anyone ever has
been at that.
Was he actually effective? Maybe not in our business, but he has a
following in some of the social sciences and that's marvelous.
He certainly made me very sensitized
to the misuse of models.

\textbf{Aldous:} And me too.

\textbf{Diaconis:} Now it's easy to just criticize modeling, but what
should we do about it?
I~wrote a paper about my version of
David's argument which was called \textit{A~place for philosophy? The rise of
modeling in statistical science}~\cite{place}.
I~tried to make a list of
what we can do.
David's approach to what we should do was
embodied in the last book he wrote~\cite{MR2489600}.
He spent years writing out with infinite clarity about
topics he had such scorn for.
I~had never quite understood why\vadjust{\goodbreak} he put so much energy into expounding
(e.g.) the Cox proportional hazards model
or the mysteries of regression.
Then he said to me, as if it were obvious, though
it hadn't occurred to me before:
``If I~say it really, really clearly, then people will see how crazy it is.''

David was a brilliant mathematician.
I~miss him daily, because we used to chat all the time.
And I~could ask him anything, from ``where to eat'' to
fine points of nonmeasurable sets.
This continued until a few years before his death.
We had written 33 papers together, and I'm a shoot-from-the-hip guy in
writing first drafts, and David was very careful,
and
very artful in his prose, and finally we got rather tired of each
other, like
an old married couple---we felt we had heard everything the other had
to say. I~found his constant negativity draining, and he found my
constant enthusiasm draining.
But we had been a pretty good pair for a long time.

Right now, Laurent Saloff-Coste and I~\cite{convol} are trying to make
a little theory of ``who needs positivity?''
What happens when you start convolving signed measures?
Infinite products are often not well-defin\-ed.
I'm sure there's some technical way of fixing that.
It's the kind of thing where David would have said, ``Let's think about it,''
and some nice math would have come out of it.
Now, with David gone, I~don't know who to ask about such things, I~don't know who cares about measure theory any more.

\textbf{Aldous:} But we all figure you have 57 collaborators, so you always
have somebody to call.

\textbf{Diaconis:}
I~do have a lot of collaborators, and that's an absolute joy, though
there's a cost. You have to own up to how little you know, and not be
afraid to make a fool of yourself.

%s6 #&#
\section{More Collaborators}
\textbf{Aldous:} Because you have had a huge number of collaborators,
we might
apologize in advance to any who are not mentioned in this conversation.
In the 1984 conversation you emphasized Martin Gardiner and Fred Mosteller
and Charles Stein and David Freedman as the people you had interacted
with and been influenced by the most by that time. Are there others
during your later career, not already mentioned, who you would like to
talk about?

\textbf{Diaconis:} Well, there's you, with exchangability and card
shuffling and mixing in MCMC, and statistics and probability in the
real world. And Laurent Saloff-Coste, an analyst who I've converted to be
somewhat of a probabilist. He was visiting Dan Stroock, and\vadjust{\goodbreak}
at that time was very far from probability, and we got into an argument,
and he was right and I~was wrong.

I've written a lot of papers with Ron Graham.
He tried to hire me when I~got my Ph.D. I~remember knocking on his
office door at Bell Labs, where he was running the math and statistics group.
I~opened the door and there was this man with a net attached to his
waist belt and going up to the ceiling. He was practicing 7 ball
juggling and the net caught dropped balls so he didn't have to pick
then up off the floor, and I~thought, this guy's great.

I've written papers with Susan Holmes, my wife, and that has its
complexity. One of the most stressful things, for each of us, is to
hear the other give a talk on our joint work. You sit there thinking,
``No, no, no, that's not the way to say it,'' and you have to keep quiet.
We've all had this experience with a graduate student, but when
it's your wife it's radically worse.
I've just finished writing a paper~\cite{shelf} with Susan and Jason
Fulman that was based on a
casino card shuffling machine that we were asked to analyze and could
in fact analyze. This was done ten years ago and the machine didn't
work, so it wasn't so polite to publish back then.

I~don't write so many papers with my graduate students---they should
get the credit for their work---but
one I~have resumed working with is Jason Fulman. I~enjoy working with
him because
he starts with a natural algebraic bent, but I~taught him to look at a
formula and look for some probability story, and he's great at it.
I~have also started writing papers with Sourav Chatterjee.
He's moving toward the probability-physics field, but I'm
encouraging him to keep some connection with statistics.

%s7 #&#
\section{Networking}
\textbf{Diaconis:}
I'm an extremely social statistician. That is, it's a lot of fun to go ask
somebody something. You need to be not too proud, to not be embarrassed
about what you don't know.
If someone asks you a question, and you don't know the answer, then
suggest someone else who might know---try to be helpful.
I~do this all the time---asking and answering, helping other people and
having them help you---but most people don't.
Learning social skills is undervalued in the research community.
There's a joy in having a community, in having people who know what
you're doing.

\textbf{Aldous:} As a related aspect of social skills,
I~tell incoming graduate students that the faculty are friend\-ly but busy;
they won't come talk to you, but you can make the effort to go talk to them.
Also, I~say to pay attention to your cohort of students---some will become
eminent in the future---and they always laugh.

\textbf{Diaconis:} Sometimes when interviewing postdocs, they think
they can come to Stanford and have \textit{you} work on \textit{their}
problem. Or they just want to work on their own thing by themselves.
It's a lot better to read some paper by the person you want to interact
with, and say, ``Can we talk about that?'', at least as a way of
getting started.
It's a simple thing to do, but most people don't do it.

%s8 #&#
\section{Old Topics Never Die}
\textbf{Aldous:}
You recently sent me an email from country X saying that most of the people
you talked to were our generation and still working on the same kind of
topics that had established their careers.
I've always liked the well-known quote from von Neumann~\cite{MR0021929}:

\begin{quote}
As a mathematical discipline travels far from its empirical source, or
still more, if it is a second and third generation only indirectly
inspired by ideas coming from ``reality'' \ldots there is a
grave danger
that the subject will develop along the line of least resistance, that the
stream, so far from its source, will separate into a multitude of
insignificant branches, and that the discipline will become a disorganized
mass of details and complexities.
\end{quote}
Of course math naturally grows in a ``one thing leads to another'' way,
but is there any test for when enough has been done on a topic and it's
time to move on?

%f1 #&#
\begin{figure*}

\includegraphics{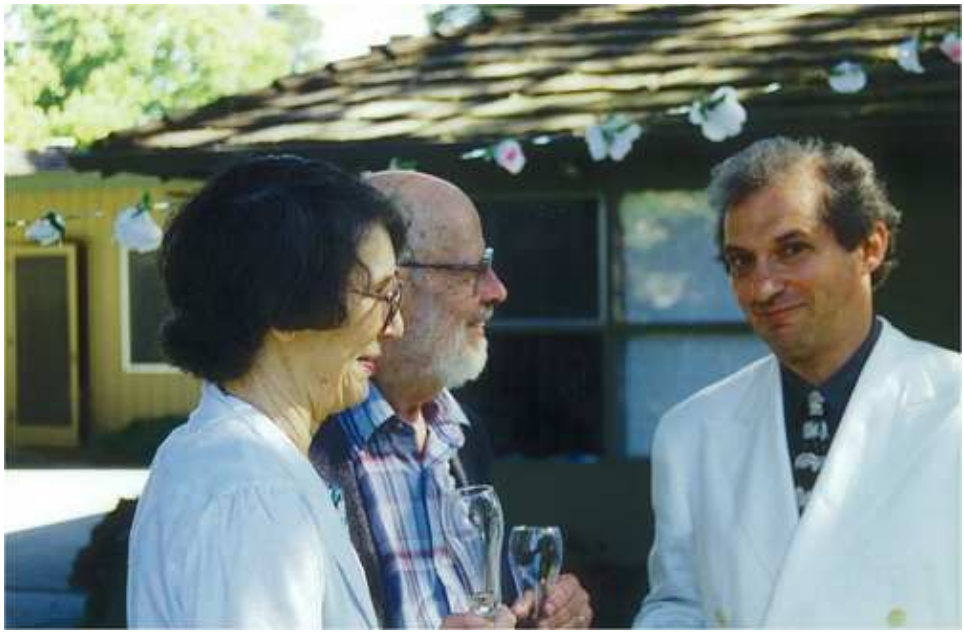}

\caption{Juliet Shaffer, Erich Lehmann, Persi Diaconis,
1997.}\label{fig1}\vspace*{-3pt}
\end{figure*}

%f2 #&#
\begin{figure*}[b]\vspace*{-3pt}

\includegraphics{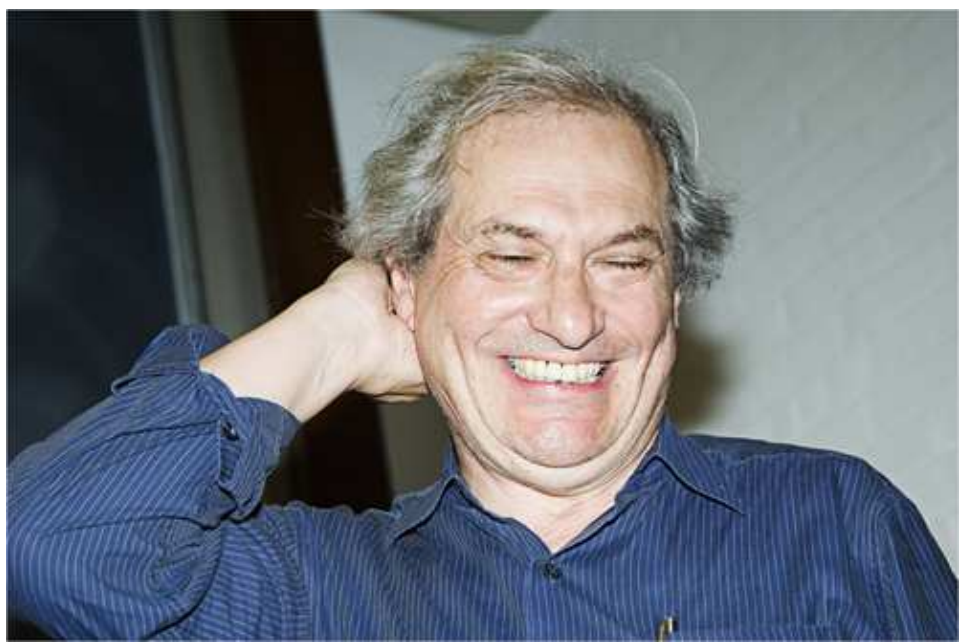}

\caption{Persi Diaconis, 2006.}\label{fig2}
\end{figure*}

\textbf{Diaconis:} It's a difficult question. Right now I'm doing some
work in
algebraic topology, a subject with enormous depth, but
many of the prominent practitioners are involved in the minutiae of how
the big machine works and don't bother to solve real problems.
They just think that if the machine is well enough developed, then it
can solve any problem that's handed to it.
I~do think it's important to try to focus on real world problems.
A lot of my motivation is MCMC, which is really used on real problems,
and, as I~said earlier,
we don't know how to give theoretical analyses of MCMC on real problems.
So what we do is problems with nice structure, say, symmetry, and hope that
will grow into something useful. von Neumann's quote is perfect---you
make a small change in a solved problem, it's still not real,
you can't do it but one of your students makes progress, and an area
grows and gets a name. It does happen that way.

Of course it's easy to criticize. One way I~try to be constructive is
take a classic like the original Metropolis algorithm applied to hard
discs in a box.
Can I~prove anything about it? I~worked very hard for five years with
wonderful analysts. We wrote papers~\cite{MR2819161,MR2504885} in the
best math journals. But our theorems are basically useless as regards
the real problem.

%f3 #&#
\begin{figure}

\includegraphics{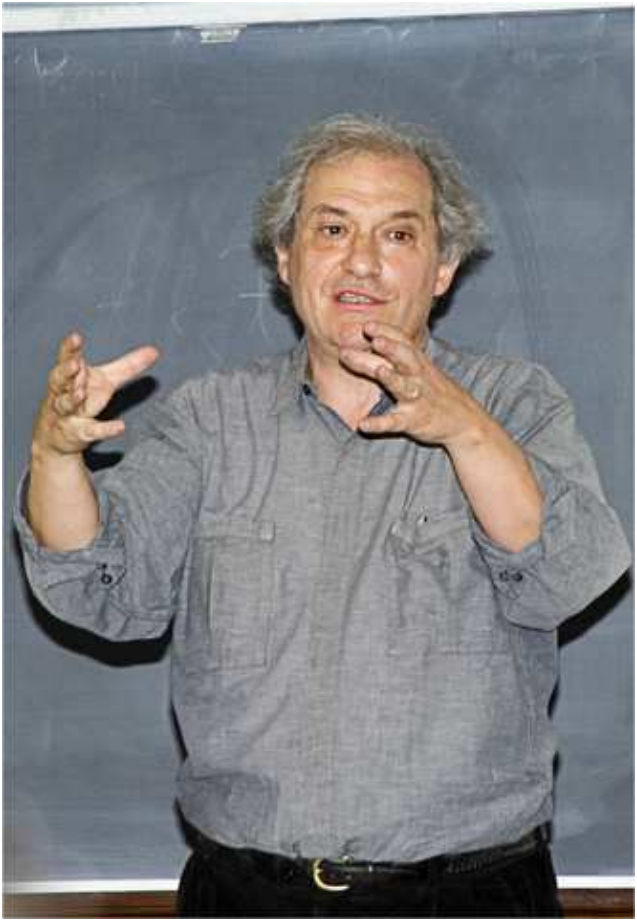}

\caption{Persi Diaconis, 2006.}\label{fig3}
\end{figure}

But again \ldots sometimes things done because they were beautiful as
pure math, then 50 years later it's just what somebody needed.
A reasonable case in point is partial exchangeability for matrices, which
David Freedman and I~were working on in 1979, and you independently
came up with a proof.
That was an esoteric corner of probability, and soon the subject went
quiet for 20 years,
but now it's completely re-emerged in contexts such as graph limits
\cite{MR2463439} and other parts of pure math~\cite{MR2827888}.
People are looking back at the old papers and
asking how did they do that.
I~just opened the \textit{Annals of Probability} and there's an article
on free probability versions
of de Finetti's theorem. Is that probability, or some other area of math?
It's very hard to know what will turn out to be useful.

\textbf{Aldous:} An unconventional idea for a workshop would be to
invite senior
people to talk about one nonrecent idea of theirs which has not been
developed or followed up by others, but which (the speaker thinks) should
be. Following Hammersley~\cite{MR0405665}, one might call these
``ungerminated seedlings of re-\break search.''~Do you have any ideas in this category?

\textbf{Diaconis:}
There's a problem that I~worked on as part of my thesis but have never
managed to get anyone else interested in. It's about summability.
A~sequence of real numbers that doesn't converge in the usual sense may be
Abel or Cesaro summable. And there are theorems that say if a sequence
is summable in scheme A, then it's summable in scheme B. I~noticed that
any time there was such a known theorem, there was a probabilistic
identity which said that the stronger method was an average of the
weaker method.
So is there a kind of meta-theorem that says this is always true?

I~once gave the Hardy Memorial Lecture at Cambridge and
wrote a paper~\cite{MR1897417} titled
\textit{G.~H. Hardy and Probability ???} with the three question marks.
Hardy notoriously didn't have much regard for applied math of any sort,
and probability was particularly low on his list. He hated being
remembered for the Hardy--Weinberg principle.
I~knew Paul Erd{\H{o}}s well, and he said that Hardy and Littlewood
were great mathematicians, but if they had had any knowledge of
probability at all, then they would have been able to prove the law of
the iterated logarithm.
That they certainly had the techniques but because they just couldn't
think probabilistically their work on that particular problem was second-rate.
Anyway, in the lecture I~wove together such stories and my own open
problems about Tauberian theory.\looseness=1

\textbf{Aldous:} Outside academia you are perhaps best known for magic
and for the ``7 shuffles suffice'' result from your 1992 paper with
Dave Bayer~\cite{MR1161056}.
I'm sure that features in every other interview you've done, so I~won't
ask again here.
More recent work of yours that attracted
popular interest was the 2007 \textit{Dynamical bias in the coin toss} paper
\cite{MR2327054}, asserting (by a mixture of Newtonian physics and experimental
observation of the initial parameters when real people performed tosses)
that there was about a 50.8\% chance for a coin to land the same way up
as tossed. I~had two undergraduates actually do the 40,000 tosses
required to have a good chance of detecting this effect, but the results
were ambiguous~\cite{40Ktosses}.
Have you or other people followed up on your paper?

%f4 #&#
\begin{figure*}

\includegraphics{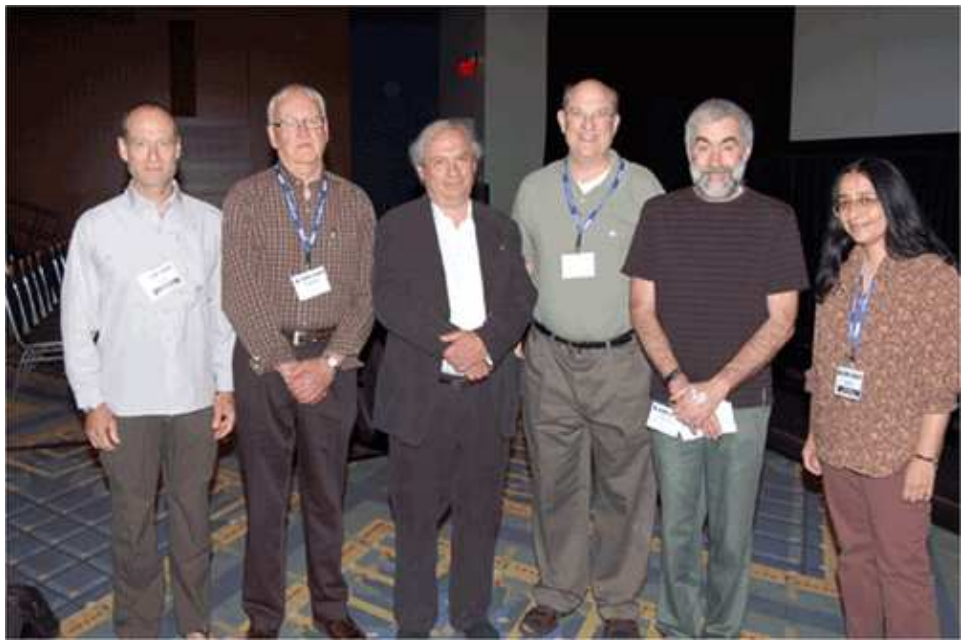}

\caption{Philip Stark, Don Ylvisaker, Persi Diaconis, Larry Brown,
Terry Speed and Ani Adhikari
at the memorial for David Freedman, 2008.}\label{fig4}\vspace*{2pt}
\end{figure*}

%f5 #&#
\begin{figure*}[b]\vspace*{2pt}

\includegraphics{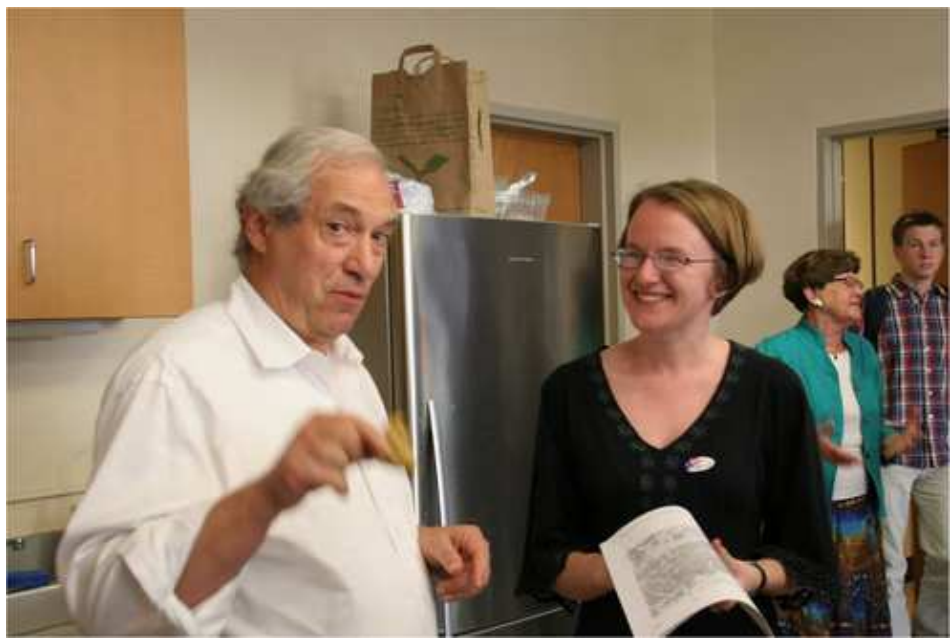}

\caption{Persi Diaconis and Elizabeth Purdom, 2010.}\label{fig5}
\end{figure*}

\textbf{Diaconis:} Aside from your students, there's a phys\-ics
group at Boston~\cite{yong2010probability} who carefully repeated our
measurements of angular velocity etc., and a
Polish group who have written a book~\cite{dynamicsgambling} on the
physics of gambling.
They reproduced our analysis and added bouncing and air resistance,
which we neglected.

Speaking of coin-tossing, every year
we get a call from ESPN and they want a two-minute spot on
``is the coin toss in the Superbowl fair?''
Of course the Superbowl coin is a big thick specially minted object,
and I~don't have much to say on that.
I~recently got a letter from a German \textit{Gymnasium} teacher
who tried to make a biased coin by making one side of balsa wood, and
he couldn't do it. I~wrote back saying that some coins are biased when
you spin them on a flat surface, but for flipping in the air
we can prove you can't make it biased \ldots

\textbf{Aldous:} \ldots by conservation of angular momentum, which a
high school physics teacher should know. You may recall that two of our
colleagues have a paper titled
\textit{You can load a die, but you can't bias a coin}~\cite{MR1963275}.

%s9 #&#
\section{Modern Times}
\textbf{Aldous:} In the 1984 conversation, when asked\break about the future
you were
wise enough not to make very specific predictions about particular topics,
but I~do notice two points. You noted there was increasing collaboration---``more and
more 2- or 3-author papers''---and we're all aware this
trend has continued. The current (October 2011) \textit{Annals of Statistics} has
only 2 out of 17 articles being single-authored, whereas going back 30
years (September 1981) it was 10 out of~17. Incidently, the total length
of the 17 articles increased from under 200 pages to almost 500 pages, a
perhaps less predictable effect. Your second point, paraphrasing
slightly, was ``I'm glad Statistics is not that kind of high-pressure field
where you have to publish every two weeks.'' But today we do have younger
colleagues who publish fifteen papers per year.

We can probably all agree that increased collaboration is A Good Thing,
but what about the increased number of papers and the implicit pressure on
young people to publish more than in our day?

\textbf{Diaconis:} Right now I'm on the hiring committees for both the
Math and Stat departments, and it's noticeable that even applicants
straight out of grad school have 3--10 papers on their CV, many of
them in pretty good journals.
How has that happened? When I~was at that stage I~just had some
technical reports. So it's just a cultural change.
We perceive an exponentially growing literature with just too many papers.
People publish the most obscure things.
But then the ability to search on the web allows us to keep track, and,
as I~said earlier,
sometimes the most obscure-looking paper turns out to contain just the
right thing.
And I~should be the last one to criticize there being too many papers,
because I'm now writing almost ten papers a year. I~would hate to have
to choose which ones I~shouldn't have written.

In our field we still referee, or pretend to referee, papers, and we
all know it can take six months or a year to get through.
I~do some work with physicists and physics is largely an unrefereed subject.
Their logic is that if somebody publishes a wrong result, the community
becomes aware of it, and then that group gets a bad reputation.
It's not that no one looks at the paper at all; someone reads the abstract
and scans the paper to check it looks reasonable. Then it gets\vadjust{\goodbreak}
published, in time maybe closer to three weeks than three months.
So our field is moving in that direction.
Publication is less and less meaningful because of the arXiv.
But as an author I~find it useful to imagine that some referee is going
to read my paper. It makes me take care about the details and the exposition.

\textbf{Aldous:} Your answer in 1984 to ``what does the future hold for
you?'' was
``just going crazy, working hard, learning more math.'' I~think we can
agree that prediction was correct. So let me ask the same question again,
and ask for your thoughts on the future of the field of Statistics, and
ask for advice to someone completing an undergraduate degree and
contemplating starting a Ph.D. program in Statistics.\looseness=-1

\textbf{Diaconis:} Yes, I~still like working hard and learning more.
Over my career
Statistics has changed so drastically it's almost unrecognizable.
Companies like Target predicting what their individual customers will
want or can be persuaded to want---this kind of
aggressive analysis of massive data sets.
So there's a lot of new Statistics for someone like me who's
classically trained. You have to find a part of it you want to learn.
For example, I'm trying to think about large networks via general
models for random graphs.
And for a theoretical statistician, looking at what applied people are
doing and asking, ``Can I~break it, can I~do it better?''
will always give us plenty to do.

About what a youngster should do \ldots for a start,
you can't learn too much about using computers.
I~lament that the academic statistics world doesn't know how to
recognize and reward
that skill appropriately. There are people who are amazing hackers and
that's an invaluable skill, but they don't get the same credit as
mathematically-focused people. I~don't know why this is, but it should change.

\textbf{Aldous:} Presumably because of the traditional ``research
$=$ papers in journals'' equation---we're so used to assessing
research contributions in that particular way.
Even though there are journals like
\textit{Journal of Computational and Graphical Statistics}, they maybe
are perceived as less prestigious.

\textbf{Diaconis:} Another piece of advice is to read classic papers.
If there's a topic that interests you, look back at what the people who
invented it actually wrote. It gives you a more concrete sense of why
they invented it and what it's about, compared to reading textbooks.
Nowadays people don't pay enough attention to such things---instead it's
``Let's try it out and write a quick paper.''

Statistics is as healthy as it's ever been.
One can see the prominence of\vadjust{\goodbreak} machine learning, but
they are really just using ideas that were developed in Statistics
twenty or fifty years ago. They are applying them---that's great---but
we are
inventing the ideas that will be applied in the next twenty or fifty years.
Statistics is a great field to be part of, and I'm still excited by it.

\section*{Acknowledgments}
I~thank Raazesh Sainudiin for proofreading
a first draft.

% imsref loaded by akundreckaite, 2012-09-28 12:55:23
%

%suskaldyti doi

\end{document}